\documentclass[aps,tightenlines,nofootinbib,prd]{revtex4}
\pdfoutput=1
\usepackage{amsmath,amscd,amssymb}
\usepackage{latexsym}
\usepackage{graphicx}
\usepackage{diagbox}

\newcommand{\imu}{{\rm i}}
\newcommand{\zr}[1]{\mbox{\hspace*{#1em}}}
\newcommand{\ID}{\mbox{{\sf 1}\zr{-0.16}\rule{0.04em}{1.55ex}\zr{0.1}}}

\begin{document}

\title{Examples for BPS solitons destabilized by quantum effects}

\author{Willem J. Meyer$^{a)}$, Herbert Weigel$^{a)}$}

\affiliation{%
$^{a)}$Institute of Theoretical Physics, Physics Department, Stellenbosch University,
Matieland 7602, South Africa}

\begin{abstract}
We investigate serval models for two scalar fields in one space dimension
with topologically stable solitons that are constructed from BPS equations. The
asymptotic behavior of these solitons fully determines their classical energies.
A particular feature of the considered models is that there are several translationally
invariant ground states that we call primary and secondary vacua. The former are those
that are asymptotically assumed by the solitons. Solitons that occupy a secondary
vacuum in finite but eventually large portions of space are classically degenerate.
Thus the quantum contributions to the energies are decisive for the energetically
favored soliton. While some of these solitons were constructed previously, we, for
the first time, compute the leading (one-loop) quantum contribution their energies.
In all cases considered we find that this contribution is not bounded from below
and that it is the more negative the larger the region is in which the soliton
approaches a secondary vacuum. This corroborates the conjecture, earlier inferred
from the Shifman-Voloshin soliton, that the availability of secondary vacua destabilizes
these solitons on the quantum level.
\end{abstract}

\maketitle

\section{Introduction}
\label{sec:intro}
Even though strictly speaking they are solitary waves, we use the notion 
{\it solitons}~\cite{Ra82,Manton:2004tk,Vachaspati:2006zz,Weinberg:2012} for 
solutions to non-linear field equations that have localized (classical)
energy densities and finite total energies. Examples for solitons in physics 
are Skyrmions~\cite{Skyrme:1961vq,Zahed:1986qz}, monopoles in $D=3+1$ space-time 
dimensions \cite{tHooft:1974kcl,Polyakov:1974ek} and vortices, strings and lumps 
in $D= 2+1$ \cite{Vachaspati:1992fi,Achucarro:1999it,Nambu:1977ag}. Solitons have almost 
uncountable applications in various disciplines: in cosmology~\cite{Vilenkin:2000jqa}, 
condensed matter physics~\cite{Schollwock:2004aa,Nagasoa:2013}, magnetic 
systems~\cite{Kosevich1990abc} as well as hadron~\cite{Weigel:2008zz} and nuclear 
physics~\cite{Feist:2012ps}.  We point to those textbooks and review articles for 
more details and further references.

Here we focus on particular models with two scalar fields, $\phi$ and $\chi$, in $D=1+1$
which have the particular feature of possessing two translationally invariant minimal energy
configurations that we call {\it primary} and {\it secondary} vacua. In addition these models 
have Bogomol'nyi-Prasad-Sommerfield (BPS) \cite{Bogomolny:1975de,Prasad:1975kr} constructions 
that simplify the static soliton equations to first order differential equations. Within
BPS constructions the classical energies are fully determined by the solitons' asymptotic 
values \cite{Adam:2016ipc}. While the solitons assume primary vacua at spatial infinity and 
are kink-like ($\phi$ can be smoothly modified into the hyperbolic tangent function without 
changes at spatial infinity), they may dwell in a secondary vacuum within a finite region of 
space. We will see that the solitons are classically degenerate in a single, continuous 
parameter that controls size of that region. Because of that degeneracy there is no unique 
classical soliton and the quantum contribution to the energy, even if small when compared to 
the classical energy, will decide which is the energetically favored soliton. This scenario has 
recently been investigated for the Shifman-Voloshin-soliton \cite{Shifman:1997wg}. It was found 
that, as a function of that single parameter, the leading, one-loop quantum correction to the energy, 
the so-called vacuum polarization energy (VPE) did not have a lower bound \cite{Weigel:2018jgq}. 
Hence no stable soliton exists in that model at the one-loop quantum level. It was hence conjectured 
that such quantum destabilization scenarios will occur whenever the soliton has access to a 
secondary vacuum. Here we will provide further examples to corroborate that conjecture.

We will employ spectral methods \cite{Graham:2009zz} to compute the VPE. These methods have
been demonstrated to be very efficient when expressing the contribution from the continuum 
quantum fluctuations as an integral over imaginary momenta \cite{Graham:2022rqk}. In section 3
we will explain that the spectral methods yield the VPE from solving a single differential
equation for the quantum fluctuations in the soliton background. This will also reveal that these 
methods can be applied straightforwardly and are less intricate than other methods as, for example, 
the heat kernel expansion~\cite{Elizalde:1995hck} which has also been applied \cite{AlonsoIzquierdo:2003gh} 
to particular soliton solutions of the Shifman-Voloshin model. We will adopt the simple 
{\it no-tadpole} renormalization scheme that fully removes the ultraviolet divergent Feynman 
diagrams. In $D=1+1$ this scheme is equivalent to the formalism of Ref. \cite{Cahill:1976im} 
which has recently been linked to a normal-ordering description \cite{Evslin:2019xte}. For 
the prime example of the kink in the $\phi^4$ model, the no-tadpole prescription yields
\cite{Graham:2009zz} the historic Dashen-Hasslacher-Neveu result \cite{Dashen:1974cj}.

In section 2 we will discuss the properties of the BPS construction for the models 
under consideration. The relevant techniques from the spectral methods will be provided
in section~3. Section 4 contains our results, both for the construction of the solitons
as well as their VPEs. Some concluding comments are contained in section 5.

This article contains quite a number of mathematical symbols. We summarize and
describe them in table \ref{tab:symbol}.

\begin{table}[t]
\begin{tabular}{|l|l|}
\hline
symbol(s) & description \\
\hline
$x$ and $t$ & space and time coordinates, respectively\\
$\mathcal{L}$ & Lagrangian (density)\\
$\phi$, $\chi$, $\phi_i$, $\gamma$ & scalar fields\\
$\phi_s$ and $\chi_s$ & secondary vacua of the scalar fields\\
$W$ & super-potential for the scalar fields\\
$\phi_i^{(s)}$ & soliton profiles of scalar fields\\
$\eta_i$ & fluctuations about the soliton\\
$M_{ij}={\rm diag}(m_1^2,m_2^2)$ & mass matrix, containing the (squared) masses of the fluctuations\\
$m_\phi$ and $m_\chi$ & masses of the fluctuations with the mapping $m_1={\rm min}(m_\phi,m_\chi)$\\
$V_{ij}$ & potential matrix for the fluctuations, generated by the soliton\\ 
$\omega$ or $\omega_k$ & frequency/energy eigenvalues of the fluctuations\\
$k$ and $k_2(k)$ & wave-numbers of the fluctuations with masses $m_1$ and $m_2$, respectively\\
$t$ and $\widetilde{t}$ & analytic continuation of $k$ and $k_2$, respectively, to the imaginary axis\\
& the context clarifies whether $t$ is the time or the imaginary momentum\\
$F_k(x)$ and $Z_k(x)$ & Jost solution and its factor function, respectively\\
$\nu(t)$ and $\nu_1(t)$ & Jost function and its Born approximation\\
$E_{\rm cl}$ & classical energy of the soliton\\
$E_{\rm VPE}$ & vacuum polarization energy (VPE)\\
$E_{\rm CT}$ & counterterm contribution to the VPE\\
$\mathcal{F}_{S,A}$, $\mathcal{F}_{\pm}$ & matrices of Jost functions in various parity channels\\
$\mu_i$ and $b$ & model parameters\\
$a$ & variational parameter for soliton profiles, defined by $\chi(0)=a\chi_s$\\
$E_0$ and $E_1$ & fit parameters for the VPE\\
\hline
\end{tabular}
\caption{\label{tab:symbol}Mathematical symbols used in the text and their meaning.}
\end{table}

\section{General structure of the sample models}
\label{sec:model}

The models that we will discuss are defined by so-called {\it super-potentials} 
$W(\phi,\chi)$ \cite{Bazeia:1996cgk} whose derivatives specify the Lagrangian
\begin{equation}
\mathcal{L}=\frac{1}{2}\left\{\dot{\phi}^2+\dot{\chi}^2 -\phi^{\prime2}-\chi^{\prime2}
-\left(\frac{\partial W}{\partial\phi}\right)^2-\left(\frac{\partial W}{\partial\chi}\right)^2\right\}\,.
\label{eq:lag}\end{equation}
Here dots and primes denote time ($t$) and space ($x$) derivatives, respectively. For 
static configurations, the classical energy then becomes a BPS construction
\begin{align}
E_{\rm cl}&=\frac{1}{2}\int dx\,\left\{\phi^{\prime2}+\chi^{\prime2}
+\left(\frac{\partial W}{\partial\phi}\right)^2
+\left(\frac{\partial W}{\partial\chi}\right)^2\right\}\cr
&=\frac{1}{2}\int dx\,\left\{\left[\phi^\prime\mp\frac{\partial W}{\partial\phi}\right]^2
+\left[\chi^\prime\mp\frac{\partial W}{\partial\chi}\right]^2\right\}
\pm W(\phi,\chi)\Big|_{x=-\infty}^\infty\,,
\label{eq:ecl}\end{align}
where the signs must be chosen such that the boundary contribution is non-negative.
Once the boundary values of the fields at positive and negative spatial infinity 
are prescribed, $E_{\rm cl}$ is minimized by the solutions to the first order equations
\begin{equation}
\phi^\prime=\pm\frac{\partial W}{\partial\phi}
\qquad{\rm and}\qquad
\chi^\prime=\pm\frac{\partial W}{\partial\chi}\,.
\label{eq:BPS}\end{equation}
As a consequence 
$E_{\rm cl}=\left|W(\phi,\chi)\big|_{x=\infty}-W(\phi,\chi)\big|_{x=-\infty}\right|$ 
whatever the solutions to the above equations look like.

Typically the super-potential is an even function of $\chi$ but it is odd in $\phi$. 
Then $\chi^\prime$ vanishes whenever $\phi$ does. In turn $\phi$ and $\chi$ will, respectively, 
be odd and even functions of the spatial coordinate when the coordinate system is defined
such that $\phi(0)=0$. Hence the only undetermined initial condition for the BPS equations 
(\ref{eq:BPS}) is $\chi(0)$. Without loss of generality we may take $\phi(-\infty)>\phi(\infty)$ 
and $W(\phi,\chi)\Big|_{x=\infty}>W(\phi,\chi)\Big|_{x=-\infty}$ so that the upper 
sign applies in Eqs. (\ref{eq:ecl}) and (\ref{eq:BPS}).
Requiring that $\phi$ is a monotonous function (like the $\phi^4$-kink) then constrains
$\frac{\partial W}{\partial\phi}\Big|_{x=0}<0$. Depending on the model parameters this
condition can be fulfilled for a range of values $\chi(0)$. If so, the classical energy
will be degenerate in $\chi(0)$ and the quantum correction to the energy will decide on
the favored configuration. 

Furthermore we have to specify the notion of primary and secondary vacua that correspond
to the translationally invariant solutions of the BPS equations (\ref{eq:BPS}). With the 
above mentioned properties of the super-potential under sign changes of the fields, 
$\frac{\partial W}{\partial\chi}=0$ when $\chi\equiv0$ so that 
$\frac{\partial W}{\partial\phi}\big|_{\chi=0}=0$ has (at least) two solutions\footnote{The
two solutions must have opposite signs which allowed us to set $\phi(0)=0$ above.} for
$\phi$. These are primary vacua and the soliton assumes one of them at negative 
spatial infinity and another one at the other end of the universe. As already mentioned,
$\frac{\partial W}{\partial\chi}=0$ when $\phi=0$. We can thus construct an alternative 
solution to $\phi^\prime=\chi^\prime=0$ by determining a constant, non-zero $\chi$
from $\frac{\partial W}{\partial\phi}\big|_{\phi=0}=0$. This is also a zero
energy configuration which we call the secondary vacuum because we cannot build a
(topological) soliton on top of it. The reason being that the profiles at positive
and negative spatial infinity would be equal since $\chi$ is invariant under spatial
reflection. However, by varying $\chi(0)$ we can find solitons on top of primary vacua
that approach such a secondary vacuum configuration within a finite but eventually 
large portion of space. It is exactly this scenario that we will investigate for a number 
of models. We will also encounter situations in which the secondary vacuum has both 
$\chi\ne0$ and $\phi\ne0$.

\section{Vacuum polarization energy (VPE)}
\label{sec:vpe}
Here we briefly describe the computation of the VPE, {\it i.e.} the leading quantum 
contribution to the total energy, in the framework of spectral methods for the case of 
two scalar fields in $D=1+1$. These methods are by now well-established and in particular 
their almost effortless use when continuing the momenta of the scattering modes to 
the complex plane has recently been reviewed \cite{Graham:2022rqk}.

The point of departure is to parameterize the two fields as
\begin{equation}
\phi_i(x,t)=\phi_i^{(s)}(x)+\eta_i(x){\rm e}^{-\imu \omega t}
\qquad {\rm for}\quad i=1,2\,.
\label{eq:lin1}\end{equation}
The index $i=1,2$ stands for either the $\phi$ or $\chi$ fields. The superscript refers 
to the soliton solutions constructed above and we have omitted the frequency argument 
for the small amplitude fluctuations $\eta_i(x)$. They are subject to the second order
differential equations
\begin{equation}
\eta_i^{\prime\prime}=-\omega^2\eta_i+M_{ij}\eta_j+V_{ij}\eta_j\,,
\label{eq:lin2}\end{equation}
that emerge from substituting the parameterization, Eq.~(\ref{eq:lin1}) into the 
Euler-Lagrangian equations derived from the Lagrangian $\mathcal{L}$, Eq.~(\ref{eq:lag}) 
and linearizing in $\eta_i$. While $M=\left(M_{ij}\right)={\rm diag}(m_1^2,m_2^2)$ contains
the mass parameters, the potential matrix, 
\begin{equation}
V_{ij}=V_{ij}(x)=\frac{1}{2}\frac{\partial^2}{\partial\phi_i\partial\phi_j}
\left[\left(\frac{\partial W}{\partial\phi_1}\right)^2
+\left(\frac{\partial W}{\partial\phi_2}\right)^2\right]_{\phi_i=\phi_i^{(s)}(x)}
-M_{ij}\,,
\label{eq:scatpot}\end{equation}
is generated by the soliton profiles and thus is time-independent so that different
frequency modes decouple. The mass parameters are determined from the model parameters 
such that plane waves solve Eq.~(\ref{eq:lin2}) for $V\equiv0$, which follows for a primary 
vacuum configuration. Once $m_\phi$ and $m_\chi$ are determined, we map the fields $\phi$ 
and $\chi$ onto $\phi_1$ and $\phi_2$ such that $m_1\le m_2$. We will provide more details 
on the matrices $M$ and $V$ in the context of the particular models in section~\ref{sec:models}.

The VPE is computed as the renormalized sum of the shift in the zero-point energies
\begin{equation}
E_{\rm VPE}=\frac{1}{2}\sum_k \left[\omega_k-\omega_k^{(0)}\right]+E_{\rm CT}\,.
\label{eq:defVPE}\end{equation}
The $\omega_k$ are the energy eigenvalues in Eq.~(\ref{eq:lin2}) and $\omega_k^{(0)}$ 
are their counterparts for $V\equiv0$. The counterterm contribution, $E_{\rm CT}$
emerges from substituting the soliton profiles into the counterterm
Lagrangian that cancels the ultra-violet divergences in the sum over the eigenmodes.
In $D=1+1$ we only need a single counterterm and we implement the {\it no-tadpole} 
condition to fix its finite piece. This condition removes the $\mathcal{O}(V)$
contribution from the sum. We can indeed remove this contribution fully and not
only at a particular renormalization scale because the only ultra-violet divergent
Feynman diagram has a loop that starts and ends at the same point and thus does
not depend on the momenta of the external line(s). We will compute the sum in
Eq.~(\ref{eq:defVPE}) from scattering data extracted from the solutions to 
Eq.~(\ref{eq:lin2}). The Born approximation to these data is $\mathcal{O}(V)$.
Hence with the {\it no-tadpole} condition we merely subtract the Born approximation
from the scattering data expression for the sum in Eq.~(\ref{eq:defVPE}) and
then drop $E_{\rm CT}$ completely.

The sum, Eq.~(\ref{eq:defVPE}) has contributions from discrete bound states
$\omega_k\le m_1$ and continuous scattering states $\omega_k>m_1$. For the considered
scenarios there are typically two zero mode bound states. One for the translational 
symmetry and one that arises because the classical energy is invariant when $\chi(0)$ 
is varied. There may be further bound states depending on the peculiarities of
the models. The continuum contribution is computed by integrating the 
dispersion relation $\omega=\sqrt{k^2+m^2_1}$ weighted by the change of the
density of states induced by the soliton, 
$\delta\rho(k)=\frac{1}{\pi}\frac{d\delta(k)}{dk}$ \cite{Dashen:1974cj,Faulkner:1977aa}
where the phase shift is extracted from the scattering matrix:
$\delta(k)=\frac{1}{2\imu}{\rm ln}\,{\rm det}\,S(k)$. However, for numerical
purpose, the phase shift is not computed from the scattering matrix 
but rather we use that the phase shift is the phase of the Jost function which 
by itself is extracted from solutions to Eq. (\ref{eq:lin2}) that resemble out-going 
plane waves as $x\to+\infty$. This function is analytic in the upper half of the 
complex momentum plane and has zeros at the imaginary wave-numbers of the bound 
states \cite{Newton:1982qc}. With the subtraction of its Born approximation we 
can then compute the integral over the continuum modes as a contour integral 
because there is no contribution from the semi-circle at $|k|\to\infty$. 
Furthermore the above mentioned zeros produce poles in the logarithmic derivative 
of the phase of the Jost function and cancel the explicit bound state contribution 
in Eq.~(\ref{eq:defVPE}). All what is left in the contour integral
is the contribution from circling the branch cut in $\sqrt{k^2+m_1^2}$ along the 
imaginary axis for $k=\imu t\pm\epsilon$ and $t\ge m_1$. After a final integration
by parts we are left with a very compact formula for the VPE
\begin{equation}
E_{\rm VPE}=\frac{1}{2\pi}\int_{m_1}^\infty\frac{dt}{\sqrt{t^2-m_1^2}}
\left[\nu(t)-\nu_1(t)\right]
=\int_0^\infty\frac{d\tau}{2\pi}\, 
\left[\nu(t)-\nu_1(t)\right]_{t=\sqrt{\tau^2+m_1^2}}\,,
\label{eq:master}\end{equation}
where $\nu$ is logarithm of the Jost function and $\nu_1$ is its Born approximation.
In either case a sum of scattering channels is understood.

Next we recall the calculation of the Jost function for theories with two
fields in $D=1+1$ which has been developed in Ref. \cite{Weigel:2017kgy} for 
the case that there is some kind of symmetry that relates the wave-equation
for $x\ge0$ and $x\le0$.
Starting point is the Jost solution that we elevate to a matrix function
\begin{equation}
F_k(x)=Z_k(x)\begin{pmatrix} {\rm e}^{\imu kx} & 0 \cr 0 & {\rm e}^{\imu k_2x}\end{pmatrix}
\qquad {\rm with}\qquad
k_2=k_2(k)\equiv k\sqrt{1-\frac{m_2^2-m_1^2}{\left[k+\imu0^{+}\right]^2}}\,,
\label{eq:JostMatrix1} \end{equation}
and $\lim_{x\to\infty}Z_k(x)=\ID$. The elements within a given column represent 
the two fields while the two columns represent the scattering channels, {\it i.e.}
out-going plane waves for only either the $\phi_1$ or $\phi_2$ fields.
The $\imu0^{+}$ prescription for the momentum of the heavier field ensures 
that $F_k(x)$ is also analytic in the gap $0\le k\le\sqrt{m_2^2-m_1^2}$.
Yet, this problem does not occur when, as motivated above, the VPE is 
computed from imaginary momenta\footnote{See Ref. \cite{Petersen:2024krb}
for the real momenta version.}.
The scattering wave-function, $\Psi_{\rm sc.(x)}$, is a linear combination of 
$F_k(x)$ and $F_{-k}(x)$.  Assuming, for the time being, that the symmetric (S) and 
anti-symmetric (A) channels decouple, the corresponding scattering matrices are 
obtained from $\Psi^{(S)\,\prime}_{\rm sc.}(0)=0$ and $\Psi^{(A)}_{\rm sc.}(0)=0$.
That is, the scattering matrices $S^{(S,A)}(k)$ are determined from $F_k(0)$ and 
$F^\prime_k(0)$. The phase shifts 
$\delta_{S,A}(k)=\frac{1}{2\imu}{\rm ln}\,{\rm det}\left[S_{S,A}(k)\right]$
are then expressed in terms of  $F_k(0)$ and $F^\prime_k(0)$. We will not present
these expression but rather turn to the imaginary axis formulation and write
$Z_{\imu t}(x)=Z(t,x)$ for the analytically continuated factor matrix, whose 
second order differential equation is
\begin{equation}
Z^{\prime\prime}(t,x)=2Z^\prime(t,x)D(t)+\left[M^2,Z(t,x)\right]+V(x)Z(t,x)
\quad {\rm with}\quad
D(t)=\begin{pmatrix} t & 0 \cr 0 & \widetilde{t}\end{pmatrix}
\label{eq:jost}\end{equation}
and~~$\widetilde{t}=\sqrt{t^2-m_1^2+m_2^2}$. The analytically continued 
combinations of $F_k(0)$ and $F^\prime_k(0)$ that enter $S_{S,A}(k)$ are
\begin{equation}
\mathcal{F}_S(t)=\lim_{x\to0}\left[Z(t,x)-Z^\prime(t,x)D^{-1}(t)\right]
\qquad {\rm and}\qquad
\mathcal{F}_A(t)=\lim_{x\to0}Z(t,x)\,.
\label{eq:defjostsym}
\end{equation}
These are the so-called Jost matrices in the symmetric and anti-symmetric 
channels. For the problem at hand the symmetric and anti-symmetric channel 
do, unfortunately, not decouple. The field potential in Eq.~(\ref{eq:lag}) 
is even in both fields. Hence the diagonal elements of the scattering potential, 
Eq.~(\ref{eq:scatpot}) are even functions of both fields while its 
off-diagonals are odd. With the above discussed properties of the profiles, it 
is then obvious that $V(-x)=\tau_3V(x)\tau_3$. Therefore the scattering problem 
decouples in the parity channels $\Psi_{\pm}(-x)=\pm\tau_3\Psi_{\pm}(x)$ and 
the relevant Jost matrices are \cite{Weigel:2017kgy}
\begin{equation}
\mathcal{F}_{\pm}(t)=\left[P_{\pm}\mathcal{F}_S(t)D_{\mp}(t)
+P_{\mp}\mathcal{F}_A(t)D_{\pm}^{-1}(t)\right]\,,
\label{eq:skjost}\end{equation}
with projectors $P_{\pm}=\frac{1}{2}\left[\ID\pm\tau_3\right]$ as well as modified 
factor matrices $D_{+}(t)=\begin{pmatrix}-t & 0 \cr 0 & 1\end{pmatrix}$ and
$D_{-}(t)=\begin{pmatrix}1 & 0 \cr 0 & -\widetilde{t}\end{pmatrix}$.
The combinations in Eq.~(\ref{eq:skjost}) essentially exchange the second (first)
rows of $\mathcal{F}_S$ and $\mathcal{F}_A$ for the positive (negative) parity channel and add 
appropriate kinematic factors. From this we find the relevant logarithm
that enters Eq.~(\ref{eq:master})
\begin{equation}
\nu(t) \equiv {\rm ln}\,{\rm det}\left[\mathcal{F}_{+}(t) \,\mathcal{F}_{-}(t)\right]\,,
\label{eq:defnu} \end{equation}
since the sum of the logarithms is the logarithm of the product. The off-diagonal 
elements of $V(x)$ only enter at second order. Hence the Born approximation is just
the sum of the Born approximations for the decoupled problem,
\begin{equation}
\nu_1(t)=\int_0^\infty dx\left[\frac{V_{11}(x)}{t}+\frac{V_{22}(x)}{\widetilde{t}}\right]\,.
\label{eq:born}\end{equation}
Now we have collected all entries to compute the VPE in the no-tadpole renormalization
scheme.

As we have summarized the theoretical background, we briefly describe the numerical 
treatment leading to the VPE as a function of $\chi(0)$. Generally, all differential 
equations are integrated with Runge-Kutta algorithms combined with adaptive step 
size controls. For a particular model described by the super-potential $W$ we 
determine the vacua, in particular the secondary vacuum $\chi_s$ of the second scalar 
field, as functions of the model parameters. Then we prescribe $0\le\chi(0)<\chi_s$ 
(and $\phi(0)=0$) to solve the first order equations~(\ref{eq:BPS}). This yields arrays 
of the profile functions $\phi_n$ and $\chi_n$ at distinct points\footnote{In all 
numerical computations '$\infty$' refers to a numerical value large enough so that the 
numerical results are stable under moderate changes of this value.} $0\le x_n<\infty$. 
These arrays enter the scattering potential $V_{ij}(\phi_n,\chi_n)$. Typically the 
adaptive step size control requires the potential at points $x\notin\{x_n\}$ and we 
interpolate between neighboring points. Using $\lim_{x\to\infty}Z_k(x)=\ID$ we then 
integrate the differential equation~(\ref{eq:jost}) to obtain $Z(t,0)$ and $Z^\prime(t,0)$.
To determine the (imaginary) momentum $t$ for which we actually integrate the differential 
equation~(\ref{eq:jost}) we define the function $\epsilon(\tau)$ subject to the 
differential equation
$$
\frac{d\epsilon(\tau)}{d\tau}
=\nu\left(\sqrt{\tau^2-m_1^2}\right)-\nu_1\left(\sqrt{\tau^2-m_1^2}\right)
\qquad{\rm with}\qquad \epsilon(0)=0\,,
$$
so that $E_{\rm VPE}=\frac{1}{2\pi}\epsilon(\infty)$. Integrating this differential 
equation with an adaptive step control is advantageous because it carefully treats 
the logarithmic behavior of $\nu(\tau)$ for $\tau\lesssim0$ that originates from the
zero modes. This behavior also requires a very tiny lower bound, $\mathcal{O}(10^{-5})$ 
when numerically integrating the above differential equation. Within a given model we 
finally obtain $E_{\rm VPE}$ as a function of the variational parameter $a$, defined
by $\chi(0)=a\chi_s$.

\section{Sample models}
\label{sec:models}

We introduce dimensionless variables and fields. This rescaling leaves an
overall factor for the Lagrangian which does not affect the classical equations
of motion. However, it enters the relation between the canonical field momenta
and their velocities. In turn it acts as a loop-counter in the sense that it 
is a relative weight between $E_{\rm cl}$ and $E_{\rm VPE}$. Fortunately, for a
given model with fixed model parameters, we will only compare the VPEs for 
configurations with equal $E_{\rm cl}$ so that this relative factor is not relevant.

In all numerical calculations we adopt sign conventions such that the kink-like 
profile, $\phi(x)$ connects a positive primary vacuum at negative spatial infinity 
to its negative counterpart at positive spatial infinity. Furthermore we take all
model parameters to be non-negative.

As a reminder we list the super-potential of the Shifman-Voloshin model \cite{Shifman:1997wg},
see also~\cite{Bazeia:1995en},
\begin{equation}
W(\phi,\chi)=\frac{1}{3}\phi^3-\phi+\frac{\mu}{2}\phi\chi^2\,.
\label{eq:SV1}\end{equation}
The primary vacua are at $\phi=\pm1$ and $\chi=0$. The secondary vacua are
at $\phi_s=0$ and $\chi_s=\pm\sqrt{\frac{2}{\mu}}$. Solutions (that are classically degenerate) 
exist for $|\chi(0)|\le\sqrt{\frac{2}{\mu}}$ and, as shown in Ref. \cite{Weigel:2018jgq},
the VPE does not have a lower bound as $|\chi(0)|\,\nearrow\,\sqrt{\frac{2}{\mu}}$ except for 
the particular case of $\mu=2$ for which the model can be mapped onto a model with two 
decoupled scalar fields that have the standard quartic interaction with spontaneous 
symmetry breaking. We next turn to three further models that exhibit similar features.

In the following we will discuss solitons in three other models with secondary vacua. 
We will encounter that the structure of these secondary vacua varies with the model
parameters. We will tabulate numerical results for our main objective, the VPEs in these 
models, as functions of the model parameters and, more enlightening, as functions of the 
variational parameters that distinguish the classically degenerate solitons. The 
dependencies of the VPEs on these variational parameters will be our central results.

\subsection{Model A}

\begin{figure}[h]
\centerline{\includegraphics[width=14.0cm,height=5.0cm]{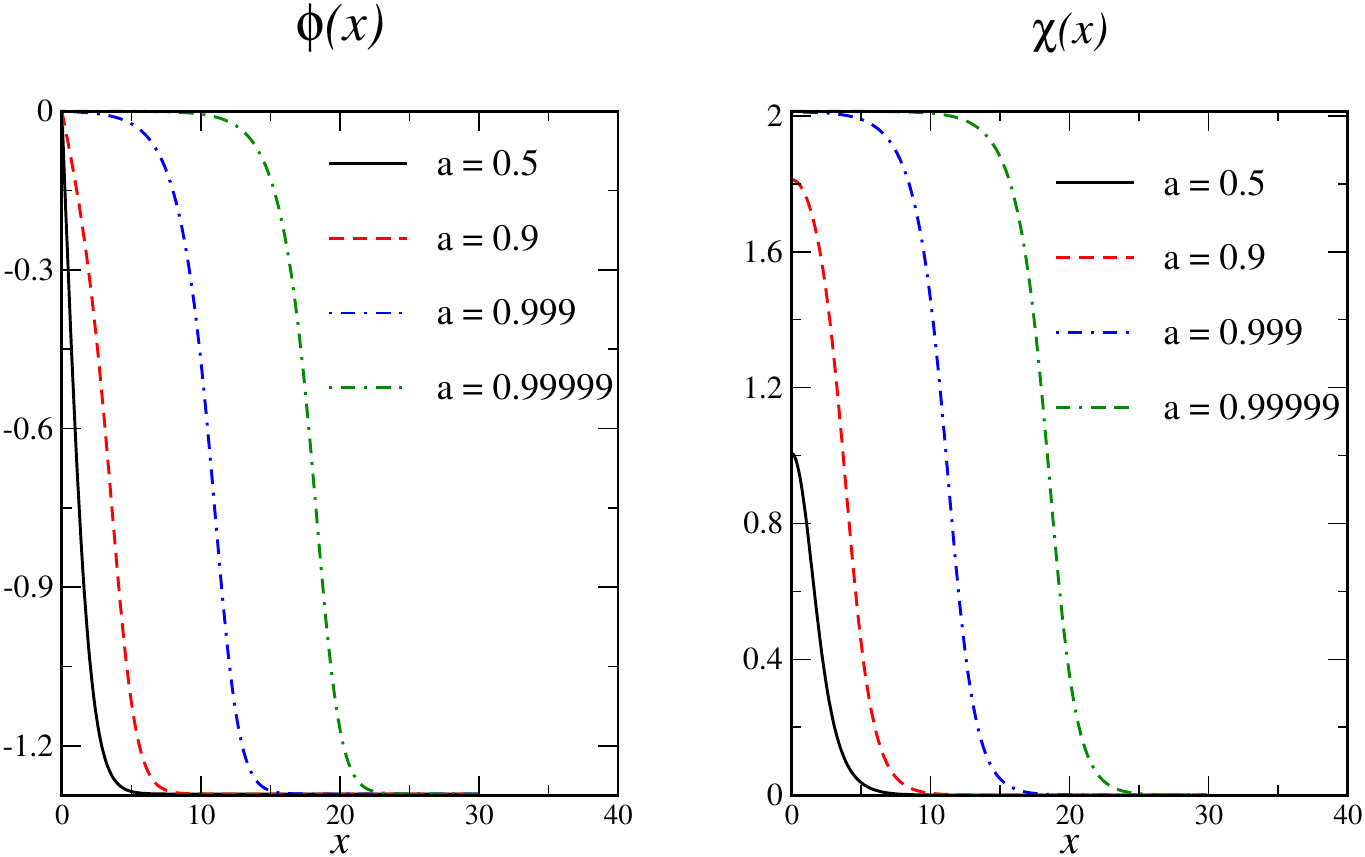}}
\caption{\label{fig:solA1}Soliton profiles for model A with $\mu_1=0.7$ and 
$\mu_2=0.6$ for various values of the variational parameter $a$. The secondary 
vacuum has $\phi_s=0$ and $\chi_s={\rm arccos}\left(1-\frac{1}{\mu_1}\right)\approx2.014$.}
\end{figure}

We replace the last term in Eq.~(\ref{eq:SV1}) by a sine-Gordon type interaction
and introduce a second model parameter
\begin{equation}
W(\phi,\chi)=\frac{\mu_2}{3}\phi^3-\phi+\mu_1\phi\left[1-\cos\chi\right]\,.
\label{eq:modelA1}\end{equation}
From the super-potential in Eq.~(\ref{eq:modelA1}) we get the static BPS equations 
\begin{equation}
\phi^\prime=\mu_2\phi^2-1+\mu_1\left[1-\cos\chi\right]
\qquad {\rm and}\qquad
\chi^\prime=\mu_1\phi\sin\chi\,.
\label{eq:modelA2}\end{equation}
The primary vacua are at $\phi=\pm\frac{1}{\sqrt{\mu_2}}$ and $\chi=0$. We expand
the fields around these vacua and read of the mass parameters 
$m_\phi=2\sqrt{\mu_2}$ and $m_\chi=\frac{\mu_1}{\sqrt{\mu_2}}$. For $\mu_1\ge\frac{1}{2}$ 
the secondary vacua are obviously at $\phi=0$ and $\chi={\rm arccos}\left(1-\frac{1}{\mu_1}\right)$ 
and the situation is similar to the Shifman-Voloshin model. Examples for the soliton profiles are shown in 
figure \ref{fig:solA1} for various values of $a=\chi(0)/{\rm arccos}\left(1-\frac{1}{\mu_1}\right)$. While, 
by construction, the profiles approach a primary vacuum at spatial infinity, they 
occupy a secondary vacuum in an ever increasing region as $a\,\nearrow\,1$. Of course,
the classical energy, $E_{\rm cl}=\frac{4}{3\sqrt{\mu_2}}$ does not depend on $a$. 

The computation of the VPE requires the potential matrix (for the case $m_\phi\le m_\chi$ so
that $\phi_1=\phi$ and $\phi_2=\chi$)
\begin{equation}
V=\begin{pmatrix}
6\mu_2\left[\mu_2\phi^2-1+\frac{\mu_1}{3}\left(1-c\right)\right]+\mu_1^2s^2
&2\mu_1\mu_2\phi s+\mu_1^2\phi sc\\[2mm]
2\mu_1\mu_2\phi s+2\mu_1^2\phi sc
&
\begin{array}{ll}
&\mu_1\left[\mu_2\phi^2-1+\mu_1\left(1-c\right)\right]c\\[0.5mm]
&\,\,+\mu_1^2s^2+\mu_1^2\phi^2\left[c^2-s^2\right]-\frac{\mu_1^2}{\mu_2}
\end{array}
\end{pmatrix}\,,
\label{eq:scatpotA}\end{equation}
with $s=\sin\chi$ and $c=\cos\chi$. By substituting the soliton profiles for $\phi$ and $\chi$
we compute the VPE with the formalism of section \ref{sec:vpe}. Not surprisingly we find that it 
does depend on the degeneracy parameter $a$. Neither are we surprised that the VPE does not have 
a lower bound as can be inferred from table \ref{tab:VPEmodelA1} in which we list the VPE for 
numerous values of $\mu_1>\frac{1}{2}$. 
\begin{table}[h]
    \centering
    \setlength{\tabcolsep}{8pt}
    \begin{tabular}{ l | c   c  c  c   c }
        \diagbox[width=50pt]{\hspace{6pt}$a$}{\hspace{6pt}$\mu_1$} 
              & 0.6 & 0.8 & 1.0 & 1.2 & 1.6 \\
        \hline
        0.0 & -0.728 & -0.843 & -0.969 & -1.107 & -1.421 \\
        0.2 & -0.731 & -0.845 & -0.971 & -1.109 & -1.423 \\ 
        0.5 & -0.757 & -0.859 & -0.982 & -1.120 & -1.441 \\
        0.9 & -0.999 & -0.961 & -1.059 & -1.206 & -1.590 \\
        0.99 & -1.548 & -1.153 & -1.190 & -1.344 & -1.832 \\ 
        0.999 & -2.116 & -1.348 & -1.321 & -1.482 & -2.073 \\ 
        0.9999 & -2.683 & -1.543 & -1.452 & -1.620 & -2.314 \\
        0.99999 & -3.250 & -1.737 & -1.582 & -1.758 & -2.555
    \end{tabular}
    \caption{\label{tab:VPEmodelA1}VPE for model A as a function of the model 
        parameter $\mu_1>\frac{1}{2}$ and the variational parameter~$a$ for $\mu_2=0.4$.}
\end{table}

\begin{figure}[h]
\centerline{\includegraphics[width=14.0cm,height=5.0cm]{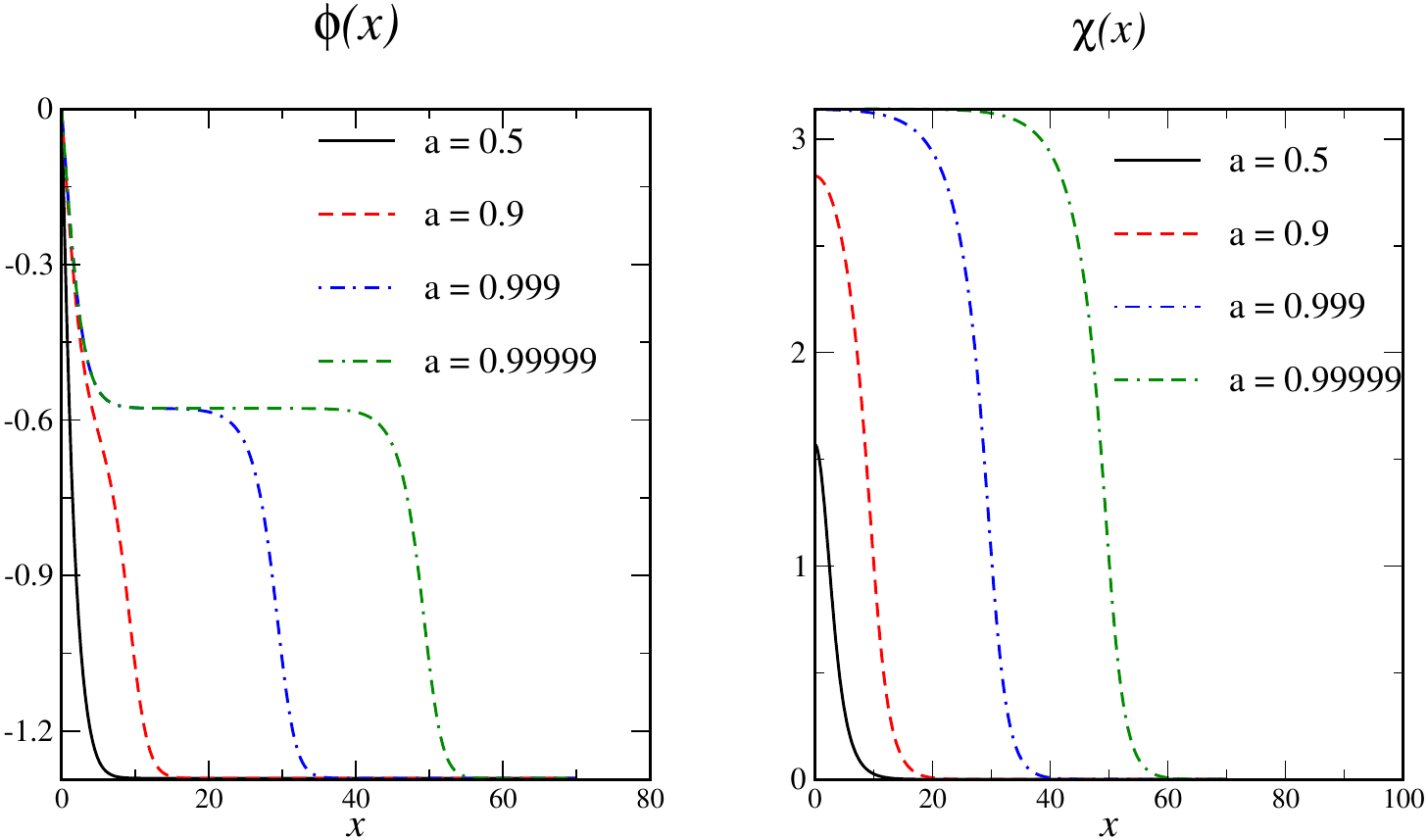}}
\caption{\label{fig:solA2}Soliton profiles for model A with $\mu_1=0.4$ and
$\mu_2=0.6$ for various values of the variational parameter $a$. Here the secondary 
vacuum is at $\phi=-\sqrt{\frac{1-2\mu_1}{\mu_2}}\approx-0.577$ and $\chi=\pi$.}
\end{figure}

For $\mu_1<\frac{1}{2}$ and $|\phi|<\frac{1}{\sqrt{\mu_2}}$ we cannot accommodate
$\phi^\prime=0$. However, there is a secondary vacuum since $\chi^\prime=0$ when
$\chi=\pi$ and then we have $\phi^\prime=0$ for $\phi=\sqrt{\frac{1-2\mu_1}{\mu_2}}$. 
Such a scenario is not contained in the Shifman-Voloshin model.
Examples for the resulting soliton profiles are shown in figure \ref{fig:solA2}
for various values of the variational parameter $a=\frac{\chi(0)}{\pi}$. 
As $a$ tends to one, the $\chi$ profile 
remains in that secondary vacuum for an ever growing region while the $\phi$ profile 
quickly drops from zero to its secondary vacuum value. Again, the VPE does not have 
a lower bound as the region grows in which the profiles assume their secondary vacua. In 
table \ref{tab:VPEmodelA2} we list the $E_{\rm VPE}$ for an exemplary set of model parameters
as a function of $a$ for the scenario that has $\chi_s=\pi$.
\begin{table}[h]
    \centering
    \setlength{\tabcolsep}{8pt}
    \begin{tabular}{ l |  c  c  c  c  c }
        \diagbox[width=48pt]{\hspace{6pt}$a$}{\hspace{6pt}$\mu_1$} & 0.10 & 0.20 & 0.30 & 0.40 & 0.45 \\
        \hline
        0.0 & -0.311 & -0.421 & -0.554 & -0.710 & -0.798 \\
        0.2 & -0.312 & -0.424 & -0.559 & -0.720 & -0.810 \\
        0.5 & -0.324 & -0.451 & -0.605 & -0.796 & -0.908 \\
        0.9 & -0.401 & -0.618 & -0.9193 & -1.429 & -1.916 \\
        0.99 & -0.439 & -0.714 & -1.147 & -2.037 & -3.122 \\
        0.999 & -0.449 & -0.750 & -1.260 & -2.405 & -3.902 \\
        0.9999 & -0.456 & -0.778 & -1.356 & -2.730 & -4.598 \\
        0.99999 & -0.463 & -0.806 & -1.449 & -3.044 & -5.271
    \end{tabular}
    \caption{\label{tab:VPEmodelA2}Same as table \ref{tab:VPEmodelA1} for various 
     model parameters $\mu_1<\frac{1}{2}$ and $\mu_2=0.1$.}
\end{table}

\begin{figure}[h]
\centerline{
\includegraphics[width=6.5cm,height=5cm]{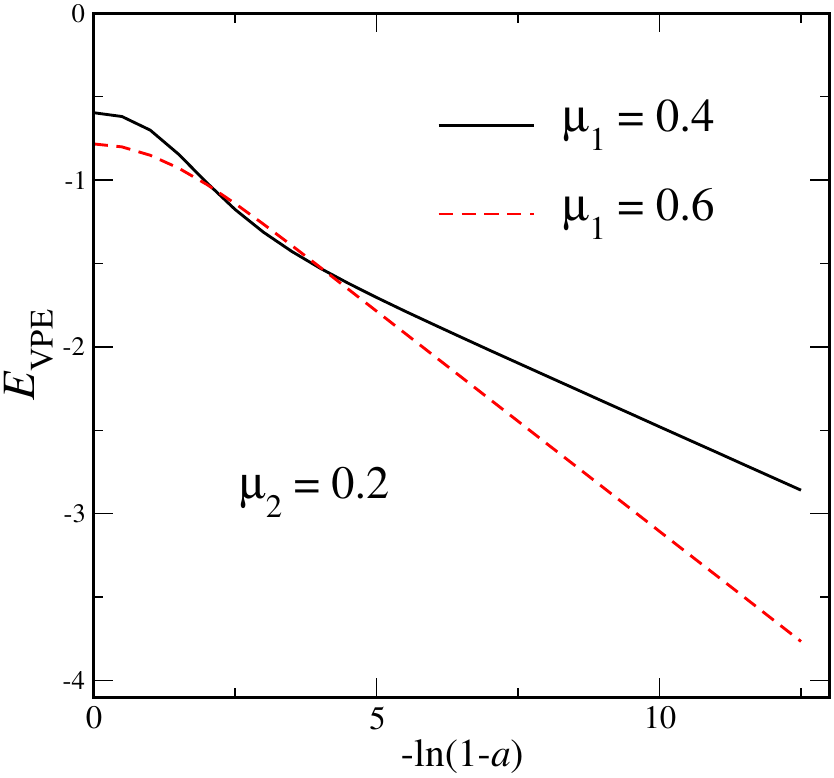}\hspace{5mm}
\includegraphics[width=6.5cm,height=5cm]{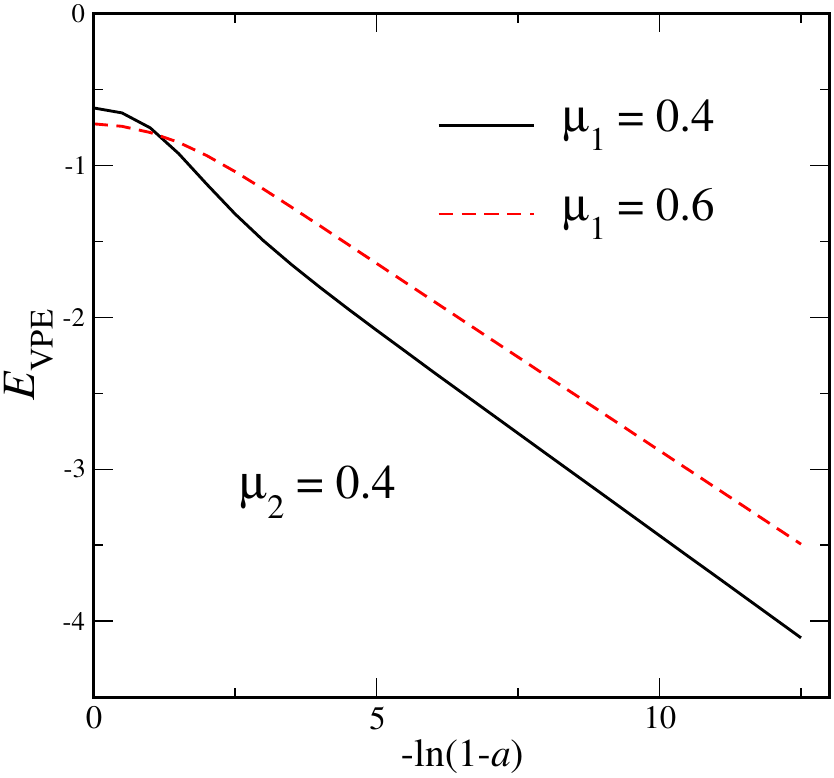}}
\caption{\label{fig:modelA_log}The VPE for the model defined in 
Eq.~(\ref{eq:modelA1}) as a function of the variable that measures the
deviation from the secondary vacuum.}
\end{figure}

In figure \ref{fig:modelA_log} we verify that as $a\nearrow1$ for both $\mu_1<\frac{1}{2}$ and
$\mu_1>\frac{1}{2}$ the VPE essentially approaches negative infinity like $\ln(1-a)$. Obviously 
there is no lower bound to the VPE and the quantum effects tend to destabilize this soliton 
whatever the secondary vacuum is.

\bigskip

\subsection{Model B}

\begin{figure}[h]
\centerline{\includegraphics[width=13.5cm,height=5.0cm]{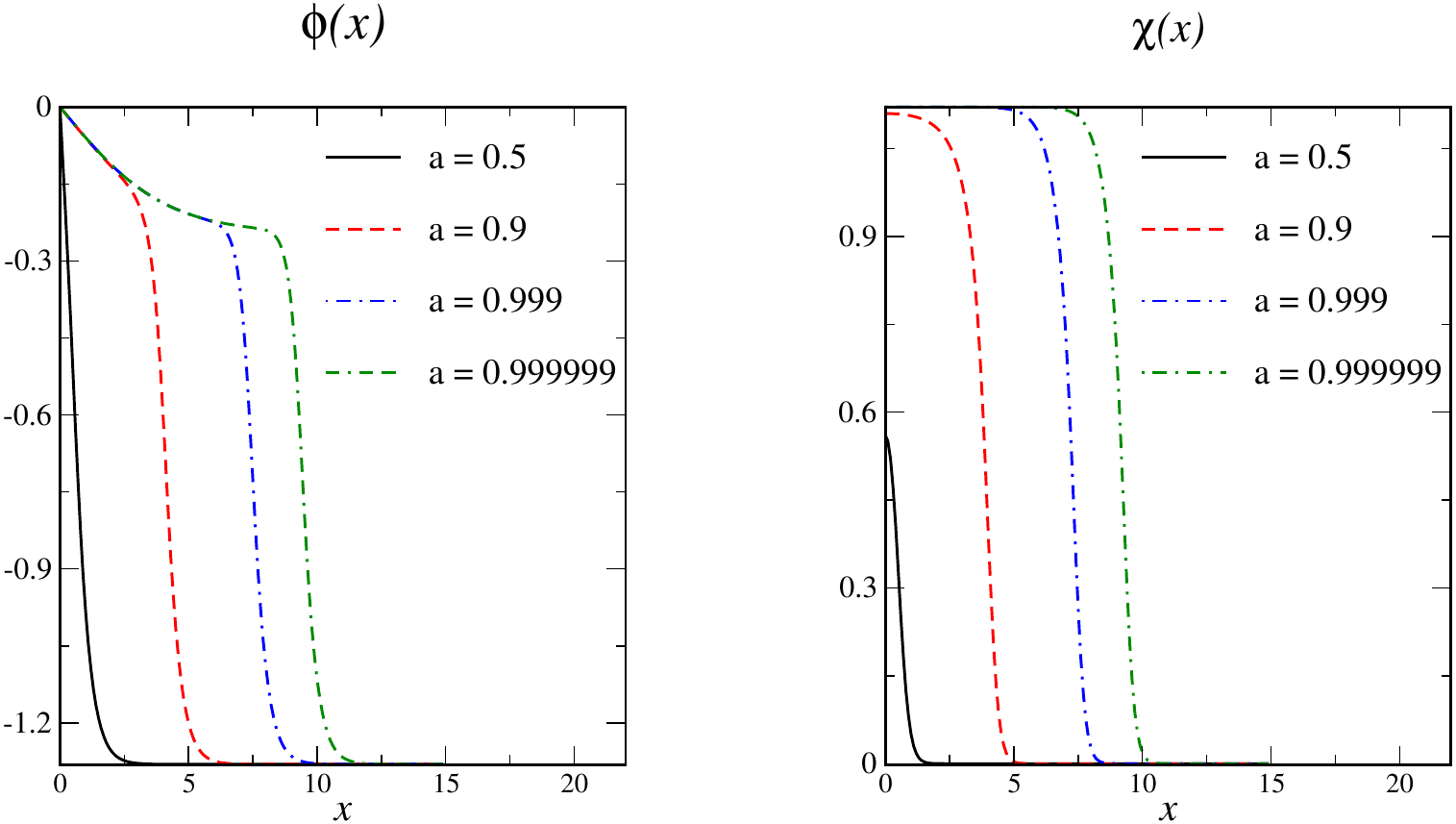}}
\caption{\label{fig:solB1}Soliton profiles for model B with $\mu_1=0.2$ and
$\mu_2=1.6$ for various values of the variational parameter $a$. The secondary
vacuum has $\phi_s=-0.2$ and $\chi_s=1$.}
\end{figure}

The next model is characterized by the super-potential
\begin{equation}
W(\phi,\chi)=\frac{1}{3}\phi^3-\phi-\phi\left(\chi^2-\mu_1\right)^2
+\mu_2\phi\chi^2\,,
\label{eq:modelB1}\end{equation}
leading to the BPS equations
\begin{equation}
\phi^\prime=\phi^2-1-\left(\chi^2-\mu_1\right)^2+\mu_2\chi^2
\qquad {\rm and}\qquad
\chi^\prime=-2\phi\chi\left(2\chi^2-2\mu_1-\mu_2\right)\,.
\label{eq:modelB2}\end{equation}
The primary vacua are at $\chi=0$ and $\phi=\pm\sqrt{1+\mu_1^2}$. The associated mass 
parameters are $m_\phi=2\sqrt{1+\mu_1^2}$ and $m_\chi=\left(2\mu_1+\mu_2\right)m_\phi$.
When $g(\mu_1,\mu_2)\equiv\mu_2\left(\mu_1+\frac{\mu_2}{4}\right)\le1$, or equivalently
$\mu_2\le2\left(\sqrt{1+\mu^2_1}-\mu_1\right)$ there are secondary vacua at 
$\chi_s^2=\mu_1+\frac{\mu_2}{2}$ and $\phi_s=\pm\sqrt{1-\mu_2\left(\mu_1+\frac{\mu_2}{4}\right)}$.
If we want to construct soliton solutions whose profiles decrease monotonously so that
$\phi(\epsilon)=-K\epsilon$ with $K>0$ and $\epsilon\,\to\,0$,
the right-hand-sides in Eq.~(\ref{eq:modelB2}) may not be positive. This requires 
$\chi^2\le\mu_1+\frac{\mu_2}{2}$. Parameterizing $\chi(0)=a\sqrt{\mu_1+\frac{\mu_2}{2}}$
the right hand side of the differential equation for $\phi$ is non-positive for $\phi\sim0$
and $a\lesssim1$ only when the above condition on the secondary vacuum is 
fulfilled. That is, the conditions on the secondary vacuum and the existence
of a soliton are identical. A sample soliton solution is shown in figure~\ref{fig:solB1}. 
The behavior generalizes: as $a\,\nearrow\,1$ the profiles approach a secondary vacuum in a 
large portion of space. However, that portion does not include $x=0$ because $\phi\ne0$ 
in that vacuum.

We abstain from presenting the potential matrix $V$ for this model and directly move to the
discussion of our numerical results for the VPE. To explore the parameter dependence efficiently
we write $\mu_2=2b\left(\sqrt{1+\mu_1^2}-\mu_1\right)$ with $b<1$. The results in terms
of the model parameter $b$ and the variational parameter $a$ are listed in 
table \ref{tab:VPEmodelB1}.
\begin{table}[h]
    \centering
    \setlength{\tabcolsep}{3pt}
    \begin{tabular}{ l | c  c  c  c c c}
        \diagbox[width=50pt]{$a$}{$b$} & 0.00 & 0.20 & 0.40 & 0.60 & 0.80 & 0.99 \\
        \hline
         0.0 & -0.940 & -1.158 & -1.403 & -1.679 & -1.990 & -2.316 \\
         0.2 & -0.940 & -1.161 & -1.409 & -1.689 & -2.002 & -2.334 \\
         0.5 & -0.947 & -1.189 & -1.458 & -1.762 & -2.109 & -2.498 \\
         0.9 & -1.086 & -1.433 & -1.823 & -2.306 & -3.035 & -5.285 \\
         0.99 & -1.289 & -1.684 & -2.096 & -2.653 & -3.761 & -13.268 \\
         0.999 & -1.462 & -1.900 & -2.265 & -2.731 & -3.883 & -17.852 \\
         0.9999 & -1.639 & -2.155 & -2.496 & -2.848 & -3.934 & -21.425 \\
         0.99999 & -1.817 & -2.425 & -2.763 & -3.009 & -3.992 & -24.677 \\
         0.999999 & -1.995 & -2.699 & -3.045 & -3.192 & -4.059  & -27.782 \\
    \end{tabular}
\caption{\label{tab:VPEmodelB1}Vacuum polarization energy for the model defined 
by the super-potential Eq.~(\ref{eq:modelB1}) for $\mu_1=0.2$. The parameters
$b$ and $a$, respectively, determine the model parameter $\mu_2$ and initial value of
the profile $\chi(0)$, {\it cf.} text.}
\end{table}
Again we can fit a function linear in $\ln(1-a)$ to the VPE for $a\lesssim1$. We find 
that the critical value for $a$ at which this behavior sets in increases with $b$. In any 
event, there is no lower bound for the VPE suggesting again that the soliton is not stable.
The effect is the more pronounced as the model parameters get closer to the limiting
case for which the soliton accesses the secondary vacuum.

\begin{figure}[h] 
\centerline{\includegraphics[width=13.5cm,height=5.0cm]{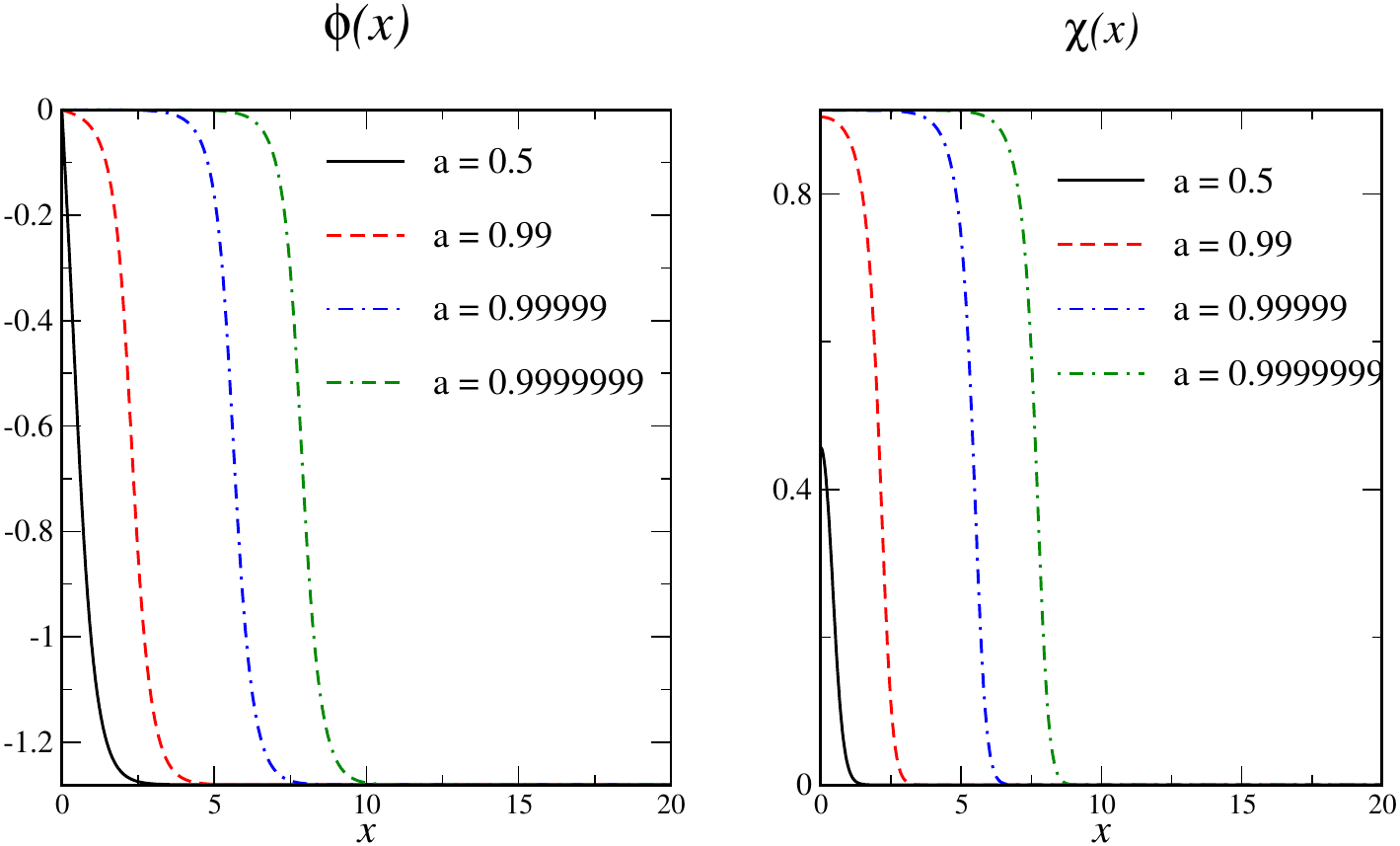}}
\caption{\label{fig:solB2}Soliton profiles for model B with $\mu_1=0.8$ and
$\mu_2=1.2$ for various values of the variational parameter $a$. The secondary
vacuum has $\phi_s=0$ and $\chi_s\approx0.91$.}
\end{figure}

As $g(\mu_1,\mu_2)\nearrow1$, $\phi$ approaches zero in the secondary vacuum. When 
$g(\mu_1,\mu_2)$ exceeds unity, $\phi_s$ stays at zero but $\chi_s^2$ decreases from 
$\mu_1+\frac{\mu_2}{2}$ to\footnote{The configuration with the positive square root 
is also a solution but does not support stable solitons.}
$\mu_1+\frac{\mu_2}{2}-\sqrt{\mu_2\left(\mu_1+\frac{\mu_2}{4}\right)-1}$ in the secondary
vacuum. A set of soliton solutions for the parameterization $\chi(0)=a\left[\mu_1+\frac{\mu_2}{2}
-\sqrt{\mu_2\left(\mu_1+\frac{\mu_2}{4}\right)-1}\right]^{\frac{1}{2}}$ and $a\nearrow1$ is shown 
in figure \ref{fig:solB2}.
The notation for the model parameters remains unchanged, but now we take $b>1$. 
\begin{table}[h]
    \centering
    \setlength{\tabcolsep}{3pt}
    \begin{tabular}{ l | c  c  c  c c c}
\diagbox[width=42pt]{$a$}{$b$~} 
& 1.10 & 1.20 & 1.30 & 1.40 & 1.50 & 1.60 \\
\hline
0.0 & -2.520 & -2.714 & -2.917 & -3.130 & -3.351 & -3.581 \\
0.2 & -2.533 & -2.726 & -2.930 & -3.142 & -3.364 & -3.595 \\
0.5 & -2.639 & -2.825 & -3.027 & -3.243 & -3.469 & -3.707 \\
0.9 & -3.536 & -3.559 & -3.720 & -3.940 & -4.195 & -4.476 \\
0.99 & -5.387 & -4.882 & -4.894 & -5.088 & -5.374 & -5.719 \\ 
0.999 & -7.318 & -6.231 & -6.080 & -6.241 & -6.555 & -6.963 \\
0.9999 & -9.251 & -7.580 & -7.265 & -7.393 & -7.736 & -8.206 \\
0.99999 & -11.184 & -8.929 & -8.450 & -8.546 & -8.917 & -9.449 \\
0.999999 & -13.117 & -10.278 & -9.635 & -9.698 & -10.097 & -10.692
    \end{tabular}
\caption{\label{tab:VPEmodelB2}Vacuum polarization energy for the model defined
by the super-potential Eq.~(\ref{eq:modelB1}) for $\mu_1=0.2$. The parameters
$b$ and $a$, respectively, determine the model parameter $\mu_2$ and 
initial value of the profile $\chi(0)$, {\it cf.} text.}
\end{table}
As before, for large enough $a$ we can approximate the VPEs listed in table \ref{tab:VPEmodelB2}
by a function that is linear in $\ln(1-a)$. 

\subsection{Model C}

\begin{figure}[h]
\centerline{\includegraphics[width=13.5cm,height=5.0cm]{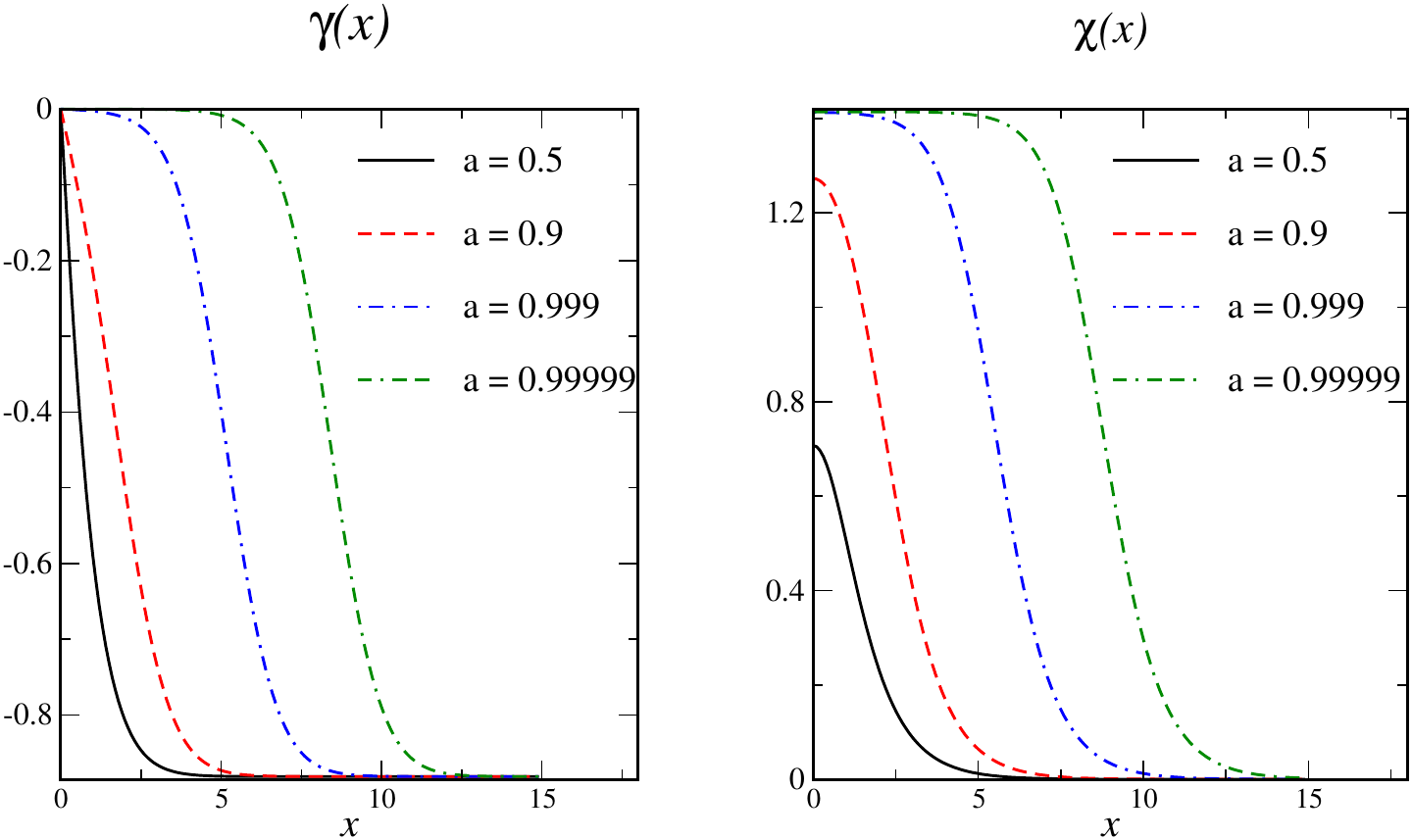}}
\caption{\label{fig:solC1}Soliton profiles $\phi=\sinh\gamma$ and $\phi$ for model C 
with $\mu=0.5$. The secondary vacuum has $\gamma_s=0$ and $\chi_s=\sqrt{2}\approx1.41$.}
\end{figure}

This model is taken from Ref. \cite{deSouzaDutra:2007vt}. It is motivated by the super-potential
of model A but subsequently the kink-type field is written as $\phi=\sinh\gamma$, so that the 
metric in the field space $\gamma-\chi$ depends on the fields. To return to a constant metric 
but nevertheless have a BPS construction the super-potential is modified to
\begin{equation}
W(\gamma,\chi)=\mu\chi^2\sinh\gamma +\int d\gamma\,\left[\sinh^2\gamma-1\right]
\left[\frac{2\mu}{\cosh\gamma}+(1-2\mu)\cosh(\gamma)\right]\,.
\label{eq:modelC1}\end{equation}
The fact that this potential is only known as an indefinite integral is not really a problem 
because we only require the $\gamma$-derivative of that integral\footnote{We would need to 
integrate from $\gamma(-\infty)$ to $\gamma(\infty)$ for the classical energy. The integral
is over $\gamma$ rather than the spatial coordinate and thus is not sensitive to the form
of the soliton.} as, for example, in the BPS equations
\begin{equation}
\gamma^\prime=\mu\chi^2\cosh\gamma+\left[\sinh^2\gamma-1\right]
\left[\frac{2\mu}{\cosh\gamma}+(1-2\mu)\cosh(\gamma)\right]
\quad{\rm and}\quad
\chi^\prime=2\mu\chi\sinh\gamma\,.
\label{eq:modelC2}\end{equation}
Again the primary vacua are at $\sinh\gamma=\pm1$ and $\chi=0$ while the secondary
vacua are at $\gamma_s=0$ and $\chi_s=\pm\frac{1}{\sqrt{\mu}}$. We therefore introduce 
the variational parameter $a$ in the initial condition $\chi(0)=\frac{a}{\sqrt{\mu}}$.
As long as $\chi\le\frac{1}{\sqrt{|\mu|}}$, the right-hand-side of $\gamma^\prime$ is 
negative for $\gamma=0$ and any value of $\mu$. Hence we expect degenerate solutions
for all $\mu>0$. The inequality arises from the demanding that the right-hand-side 
of $\chi^\prime$ is negative when $\chi>0$ and $\gamma<0$. Furthermore the mass parameters
$m_\gamma=4\left(1-\mu\right)$ and $m_\chi=2\mu$ are obtained from expanding 
Eq.~(\ref{eq:modelC2}) around the primary vacuum ($\sinh\gamma=\pm1, \chi=0$). Thus the 
existence of a BPS soliton on top of the above mentioned primary vacua yields the upper 
bound $\mu<1$. Profiles determined subject to this condition are shown in figure \ref{fig:solC1}.
We observe the by now familiar behavior that the soliton approaches the secondary 
vacuum as $a\,\nearrow\,1$.

Again we abstain from displaying the bulky expressions for the scattering potential and 
immediately list the numerical results for the VPE in table \ref{tab:VPEmodelC1}.
\begin{table}[h]
    \centering
    \setlength{\tabcolsep}{8pt}
    \begin{tabular}{ l | c  c  c  c  c }
        \diagbox[width=50pt]{\hspace{6pt}$a$}{\hspace{6pt}$\mu$} & 0.4 & 0.6 & 0.8 & 0.9 & 0.95 \\
        \hline
        0.0 & -1.076 & -0.899 & -1.046 & -1.482 & -2.050 \\ 
        0.2 & -1.090 & -0.908 & -1.075 & -1.541 & -2.142 \\ 
        0.5 & -1.169 & -0.958 & -1.233 & -1.859 & -2.630 \\
        0.9 & -1.487 & -1.080 & -1.683 & -2.828 & -4.152 \\
        0.99 & -1.886 & -1.126 & -1.998 & -3.658 & -5.539 \\ 
        0.999 & -2.268 & -1.159 & -2.273 & -4.414 & -6.816 \\ 
        0.9999 & -2.647 & -1.191 & -2.546 & -5.167 & -8.094 \\  
        0.99999 & -3.026 & -1.223 & -2.819 & -5.919 & -9.367
    \end{tabular}
    \caption{\label{tab:VPEmodelC1}VPE of model C for different values of the model 
parameter $\mu<1$ and the variational parameter~$a$.}
\end{table}

For $\mu>\frac{1}{2}$ and $\chi=0$ the right-hand-side of the differential equation 
for $\gamma$ has another zero at $\cosh^2\gamma=\frac{2\mu}{2\mu-1}$ or 
$\sinh^2\gamma=\frac{1}{2\mu-1}$, which is less than 1 for $\mu>1$. In that case the 
soliton assumes $\pm{\rm arcosh}\left(\sqrt{\frac{2\mu}{2\mu-1}}\right)$ at spatial 
infinity\footnote{This configuration actually is a maximum of the action density for 
$\frac{1}{2}<\mu<1$.}. The mass parameters for these primary vacua are 
$m_\gamma=\frac{4\mu-4}{\sqrt{2\mu-1}}$ and $m_\chi=\frac{2\mu}{\sqrt{2\mu-1}}$. 
The expressions for the secondary vacua are as in the case with $\mu<1$.

\begin{figure}[h]
\centerline{\includegraphics[width=13.5cm,height=5.0cm]{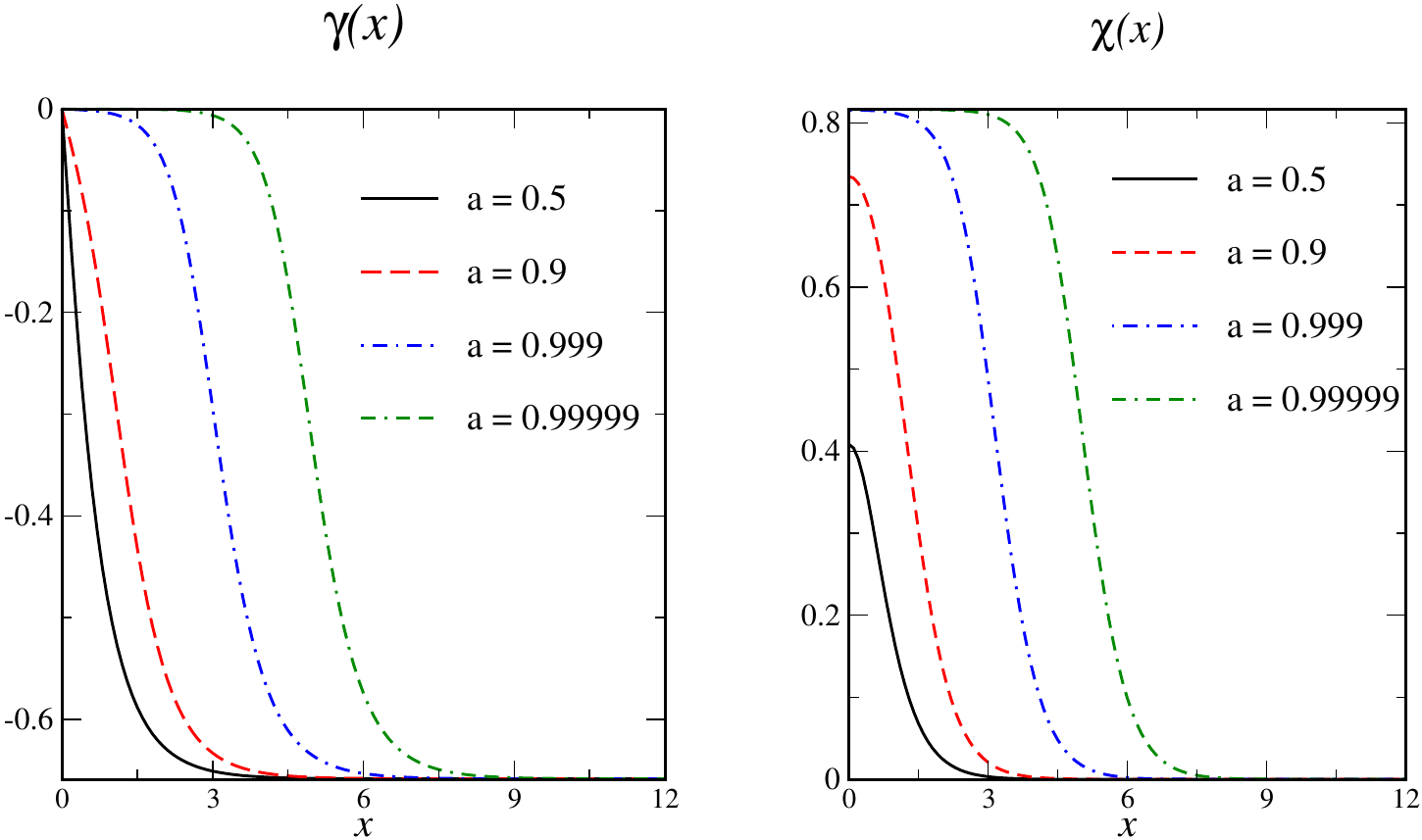}}
\caption{\label{fig:solC2}Soliton profiles $\phi=\sinh\gamma$ and $\phi$ for model C
with $\mu=1.5$. The primary vacua are at $\phi=\frac{\pm1}{\sqrt{2\mu-1}}\approx\pm0.71$
or $\gamma\approx\pm0.66$ and $\chi=0$. The secondary vacuum has $\phi_s=0$ and 
$\chi_s=\sqrt{\frac{2}{3}}\approx0.82$.}
\end{figure}

\begin{table}[h]
    \centering
    \begin{tabular}{ l | c  c  c  c  c  }
\diagbox[width=44pt]{$a$}{$\mu$} & 1.05 & 1.10 & 1.20 & 1.40 & 1.60 \\  
\hline
0.2 & -2.312 & -1.814 & -1.464 & -1.322 & -1.355 \\
0.5 & -2.830 & -2.182 & -1.700 & -1.457 & -1.448 \\  
0.9 & -4.465 & -3.342 & -2.439 & -1.865 & -1.720 \\  
0.99 & -5.996 & -4.401 & -3.082 & -2.183 & -1.908 \\   
0.999 & -7.415 & -5.379 & -3.671 & -2.469 & -2.071 \\     
0.9999 & -8.828 & -6.352 & -4.256 & -2.752 & -2.232 \\
0.99999 & -10.239 & -7.325 & -4.841 & -3.036 & -2.394
    \end{tabular}
    \caption{\label{tab:VPEmodelC2}Same as table \ref{tab:VPEmodelC1} for $\mu>1$.}
\end{table}

From the graphs in Figure~\ref{fig:solC2} and the data in Table~\ref{tab:VPEmodelC2} we
observe that the resulting VPEs show the same tendency as the other models (including
the Shifman-Voloshin case) to approach negative infinity like $\ln(1-a)$.

\subsection{A note on zero modes}

In the context of our numerical analysis we have verified the existence of 
the two zero modes mentioned in section \ref{sec:vpe}.  With the current 
techniques an effortless way to do so is to numerically determine $\alpha$ 
in $\nu(t)\approx\alpha\ln(t-m)$ for $t\,\searrow\,m$ since the Jost function 
has zeros at the imaginary momenta associated with the bound state energies. 
In almost all cases the numerical fit indeed produced results consistent with 
$\alpha=2$. The only exceptions were observed in cases with the soliton being 
in secondary vacua with $\phi\ne0$ for large regions of space. In those cases 
the fit yielded $\alpha\approx3$. We then employed techniques as in section III 
of Ref. \cite{Weigel:2017kgy} and found that the third (would-be) zero mode was 
dynamical: it emerged from an ordinary bound state with a non-zero energy 
eigenvalue in the limit $a\,\nearrow\,1$.  When the secondary vacuum does not 
include $\phi=0$, the profile has distinct plateaus along the negative and 
positive half-lines. Moving around those plateaus produces two independent zero 
modes in addition to the translational one.

\section{Conclusion}

We have investigated a number of BPS solitons in one space dimension. The models under 
consideration have several distinct translationally invariant zero energy solutions, 
at least for certain ranges of model parameters. We refer to these solutions as 
primary and secondary vacua. At negative spatial infinity topologically stable 
solitons approach a primary vacuum but its negative at the opposite end. Yet, the 
soliton may close in on a secondary vacuum in an arbitrarily large portion of 
space without any classical energy cost so that there are infinitely many classically 
degenerate solutions. We must then include the quantum corrections to identify the 
energetically favored solution. For this purpose we have computed the leading, 
one-loop quantum correction which is the vacuum polarization energy (VPE) of such 
solitons.

The sample models considered here have two scalar fields $\phi(x)$ and $\chi(x)$ with 
the corresponding static soliton profiles being respectively odd and even under spatial 
reflection. This guarantees topological stability and selects the primary vacua as those 
with $\chi\equiv0$. On the other hand the secondary vacua have $\chi\equiv\chi_s\ne0$. 
The value of $\phi_s$ in secondary vacua as well as other peculiarities of the secondary 
vacua depend on the model parameters. Further conditions on the availability of these 
vacua arise from requiring the existence of localized static solutions.

We have introduced the variational parameter $a=\frac{\chi(0)}{\chi_s}$ as the measure 
of how the spatially symmetric component of the soliton approaches its secondary vacuum, 
and have found that the classical energy is independent of $a$ by the BPS construction. 
On the other hand the VPE decreases like $E_{\rm VPE}\sim E_0-E_1\ln(1-a)$ with $E_1<0$ as 
$a\nearrow1$ and hence has no lower bound. In the same limit the soliton dwells in a secondary
vacuum for an increasing and unbound region. Though $E_0$ and $E_1$ are model dependent, 
this functional behavior has been reproduced in all of the considered models. Hence our 
computations corroborate the earlier conjecture \cite{Weigel:2018jgq} that quantum corrections 
destabilize classically degenerate BPS solitons when they may approach secondary vacua in an 
arbitrarily large portion of space.

Eventually soliton solutions with constant $\chi(x)\equiv\chi_s$ and $\phi(x)\ne0$ exist when
$\phi_s\ne0$. They connect $\phi_s$ and $-\phi_s$ along the coordinate axis. These borderline 
cases have different classical energies when $\phi_s$ is not a primary vacuum. Since these
solutions are merely variants of the ordinary kink we have not considered them here.

Of course, when the first order correction is sizable, higher orders may be as relevant. Studies 
on higher order quantum corrections for solitons have just recently commenced~\cite{Evslin:2021vgk}.

\acknowledgments{H.\@ W.\@ is supported in part by the National Research Foundation 
of South Africa (NRF) under grant~150672.}

\end{document}